# Excel: A Financial Modelling System

Peter Bartholomew Excel MVP
NAFEMS
peter.bartholomew1@btinternet.com

Luke Phillips FMI CFM
Access Analytic
lphillips@accessanalytic.com.au

**ABSTRACT**

*Since the calculation engine of Excel was the subject of a major upgrade to accommodate Dynamic Arrays in 2018, there has been a series of seismic changes to the art of financial modelling. This paper will show the traditional financial modelling techniques for spreadsheets can be replaced by radically different approaches that have significant gains in readability, performance and flexibility. With so much power added to the Excel engine in recent years, it is the responsibility of financial modellers to ensure that they are building in the simplest and most auditable way possible going forward.*

## 1. Introduction

This paper is the fifth in a sequence of papers that, somewhat by chance, have documented the significant changes that have occurred in Excel since 2018. The common theme across the series is the approach to constructing spreadsheet solutions as an exercise in programming, rather than relying on traditional tips and tricks.

The first paper was to a large degree written as a response to the FAST financial modelling standard that had been released the previous year.[FAST 2015]. The standard assumed the use of calculation blocks and required that defined Names should only be used under very restrictive circumstances and the use of array formulas should be avoided. Since the present author's experience of using Excel for engineering problems was based upon the use of defined names and array formulas, any suggestion that the new standard offered 'best practice' would sit very uncomfortably. The 2016 paper demonstrated a very different way of working that was built upon functionality that had been present, but infrequently used, within Excel since it was first ported to Windows in 1987.

A major overhaul of the Excel calculation engine led first to dynamic arrays and later to the introduction of **LET** variables, **LAMBDA** function declaration, recursion, higher-order helper functions, and array shaping functions. This has opened endless possibilities in financial modelling that were previously only possible with VBA and therefore much slower models. With VBA out of the question for most companies due to security concerns, dynamic arrays provide a way forward.

Progress has come in a number of incremental steps and at each step the ultimate goal of building financial models without accessing the traditional spreadsheet methodology has come a little closer. The 2019 paper shows dynamic arrays and spilt ranges in use but qualifies the excitement by pointing out that both the aggregation of 2D arrays by column and the accumulation of cash flows to return balances require workarounds, with options involving the **MMULT** function being shown.

By the time of the 2023 EuSpRIG Conference, both these issues had been addressed natively within Excel through the introduction of the **LAMBDA** helper functions **BYROW** and **SCAN**. **SCAN** addressed the problem of corkscrews being treated as circular references in the absence of array breakup. By 2025, the first author could assert that Excel is now capable of providing a functional programming environment that no longer relies upon the assumptions such as A1 cell referencing or filling formulas down or to the right. Since the 'modern' environment requires more time to be spent entering formulas within the AFE than on the worksheet itself, even the 'tips and tricks' that have been honed over the years to make spreadsheet navigation more efficient are largely unnecessary. The paper did address financial modelling, with the key observation being that the closing cash balance or revolver debt requires knowledge of every cash flow within the entire model, and the prior balances must be referenced through the use of **SCAN** if Excel is not to report circular reference errors.

With most financial modeller's background being in finance and accounting, a programmatic approach to modelling has been a novel, if not somewhat alien, concept so far. This paper aims to show that the novel approaches to financial modelling are on their way to becoming the new standard with more accepted, tested and repeatable logic. These novel approaches are not just leading to gains in logic efficiencies, but also computational efficiencies that can speed up financial models. We know that models that perform well, are easier to use and enable greater functionality and use, benefiting both the modeller and the companies they represent.

## 2. Contributions

This paper will be a discussion on the newer features available in Excel, and how advanced techniques can utilise these features to maximise the dependability and reliability of a financial model and the financial modelling process. A key and novel approach has been developed. This approach utilises a Central Routing Layer (CRL), and an approach to financial modelling with the same name which enables significant modelling speed, auditability and repeatability gains.

The CRL approach treats financial modelling through a data-centric lens, structuring the spreadsheet so it can operate as a reliable data source rather than a collection of manually linked calculations. Under this approach, links to the financial statements are automated through a central layer, making the process fully repeatable and consistent across statements. Users define their accounting and modelling rules up front and then apply them systematically throughout the model. This structure enables templating, standardised processes, and greater efficiency for both the modeller and the client.

Financial statements, automated checks, audit tabs, model circularity and Business Intelligence layers can all be re-configured across models rapidly, with consistent named **LAMBDA**s, as opposed to traditional methods where changes are significantly more labour intensive.

## 3. Factors changing financial modelling

### Dynamic Arrays

Whereas the previous papers addressed Excel but were not specific to any given discipline or area of application, this paper addresses issues that arise within the financial modelling domain. The introduction of dynamic arrays to Excel was the start of a real change in Excel-based financial modelling. Traditional approaches involve a fill down and right approach that require orders of magnitude more formula cells. There was a huge degree of excitement in the financial modelling community about being able to build models that have the same anchored time series, that allowed model timeframes to change on demand and calculate quicker than regular models.

In the EuSpRIG 2023 Conference, Giles Male presented the Craig Hatmaker's 5g component-based software engineering concept which remains the gold standard for the use of standard Excel **LAMBDA** functions within shared libraries. Whilst it may be possible to cover the requirements of many financial modelling problems using standard libraries, there will always be models with unsupported characteristics, and it is one such model, introduced by the second author, that we will focus on as an example in this paper.

Using dynamic arrays in an equity release model allowed the reduction of the number of formulas used to go from ~ 200,000 to 158, in one sheet[1]. It is worth noting traditional cell-based models built using FAST standard would be using consistent formula blocks, that should be acting and calculating like a dynamic array. But this type of calculation block, built using traditional methods, and especially with large 2D arrays, can cause a model to perform slowly and take an extended amount of time to update and work with. Dynamic arrays work differently; they force the user to accept the calculation block – if a user tries to overtype a part of the spill range, they will see a #SPILL error come into the model.

[1] Formula in this case is described as a cell with an "=" sign as the first character, and not the spilled section of an array.

In large corporate financial models, speed is hugely important for several reasons. Firstly, a model that responds well to changes in inputs, can switch assumptions seamlessly and measure sensitivities much quicker. Additionally, the user of the model often must respond rapidly to late changes before deadlines. For these reasons, having a slow calculating financial model can be a burden, and financial modellers should strive to make their models as efficient as possible for the end user. Craig Hatmaker has demonstrated the dynamic array financial modelling can be much faster than the traditional cell-based alternative, where he ran a 500,000-row amortisation schedule with dynamic arrays and traditional cell-based formulas. The traditional cells took 15 seconds to calculate, whilst the dynamic arrays took only 2 seconds. Whilst the effect of dynamic arrays in a financial model would likely not produce such significant speed improvements, they still represent a more efficient calculation method compared to traditional spreadsheets.

### LET( ) for Auditable logic blocks

The introduction and more popularised use of **LET**( ) in dynamic array calculations allows the user to be significantly more confident in the calculation blocks that are being built up. Let us take an example financial model in the equity release space[2]. The loan funder is interested to know what the ending balance of the loan book is going to be in each monthly tranche going into the future. This is a complex calculation, in both traditional and dynamic array modelling approaches. It requires 5 discrete blocks of logic – time series of each tranche, the initial draw, a cpi applied to the initial draw, the runoff of the loans and the compounding balance of the loans still on the book.

Whilst the ending balance of the loans will make some rational sense to a modeller or auditor, they will want to check each option. For a traditional modelling approach, each of the ~6000 cells will need to be copied over. A mistaken keyboard click can cause errors to hide silently within that calculation block. A **LET**-based dynamic array, however, just needs a single switch in the calculation to view discrete steps in the model. Each step of the logic is given a name inside the **LET** — the carryforward balance, the runoff, the reverse index, and so on — and the very last line decides which of those named steps is returned and spilled to the sheet. To check any one of them, we can change the final line[3].

Original:

```
=LET(
   _LoanPeriod,    Timing!$L$19#,
   _Carryforward,  SCANHλ($K$1176#*0, $L$1053#, LAMBDA(a,b, a+b)),
   _Runoff,        (1 - XLOOKUP(_LoanPeriod, Runoff!$L$2#, Runoff!$L$29#, 0)),
   _ReverseIndex,  XLOOKUP(_LoanPeriod, Runoff!$L$2#, Runoff!$L$31#, 0),
   _Result,        _Carryforward * _ReverseIndex * _Runoff,
   _Result)
```

[2] Equity Release Model is a loan origination business modelling sample built to be viewed with this paper

[3] Equity Release Model, Sheet "Book", cell L1176

| | 31-Jul-27 | 31-Aug-27 | 30-Sep-27 | 31-Oct-27 | 30-Nov-27 | 31-Dec-27 | 31-Jan-28 | 29-Feb-28 | 31-Mar-28 | 30-Apr-28 | 31-May-28 |
|---|---|---|---|---|---|---|---|---|---|---|---|
| 31-Jul-27 | 924,519 | 930,907 | 937,332 | 943,794 | 950,291 | 956,823 | 963,387 | 969,984 | 976,610 | 983,265 | 989,945 |
| 31-Aug-27 | - | 927,604 | 934,013 | 940,460 | 946,943 | 953,462 | 960,015 | 966,602 | 973,220 | 979,869 | 986,545 |
| 30-Sep-27 | - | - | 930,599 | 937,029 | 943,496 | 950,001 | 956,540 | 963,115 | 969,723 | 976,362 | 983,032 |
| 31-Oct-27 | - | - | - | 2,801,098 | 2,820,452 | 2,839,919 | 2,859,496 | 2,879,181 | 2,898,970 | 2,918,860 | 2,938,846 |
| 30-Nov-27 | - | - | - | - | 2,906,716 | 2,926,800 | 2,947,001 | 2,967,316 | 2,987,743 | 3,008,279 | 3,028,918 |
| 31-Dec-27 | - | - | - | - | - | 3,016,640 | 3,037,484 | 3,058,449 | 3,079,532 | 3,100,732 | 3,122,044 |
| 31-Jan-28 | - | - | - | - | - | - | 3,130,722 | 3,152,353 | 3,174,111 | 3,195,992 | 3,217,994 |
| 29-Feb-28 | - | - | - | - | - | - | - | 4,611,473 | 4,643,336 | 4,675,385 | 4,707,615 |
| 31-Mar-28 | - | - | - | - | - | - | - | - | 4,785,867 | 4,818,935 | 4,852,196 |
| 30-Apr-28 | - | - | - | - | - | - | - | - | - | 4,966,322 | 5,000,637 |
| 31-May-28 | - | - | - | - | - | - | - | - | - | - | 5,154,136 |
| 30-Jun-28 | [illegible] | [illegible] | [illegible] | [illegible] | [illegible] | [illegible] | [illegible] | [illegible] | [illegible] | [illegible] | [illegible] |

*Figure 3.1: The spilled output of the LET formula — ending loan-book balance by monthly tranche.*

In the original formula, the last line returns _Result — the ending loan-book balance we want (Figure 3.1). To audit the step that feeds it, we swap that last line for the name of the step we want to see. Here we have changed it to return _ReverseIndex instead, and the spill immediately shows that intermediate working rather than the final answer (Figure 3.2). The rest of the formula is untouched.

Changed for testing:

```
=LET(
  _LoanPeriod,    Timing!$L$19#,
  _Carryforward,  SCANHλ($K$1176#*0, $L$1053#, LAMBDA(a,b, a+b)),
  _Runoff,        (1 - XLOOKUP(_LoanPeriod, Runoff!$L$2#, Runoff!$L$29#, 0)),
  _ReverseIndex,  XLOOKUP(_LoanPeriod, Runoff!$L$2#, Runoff!$L$31#, 0),
  _Result,        _Carryforward * _ReverseIndex * _Runoff,
  _ReverseIndex)
```

| | 30-Jun-27 | 31-Jul-27 | 31-Aug-27 | 30-Sep-27 | 31-Oct-27 | 30-Nov-27 | 31-Dec-27 | 31-Jan-28 | 29-Feb-28 | 31-Mar-28 | 30-Apr-28 | 31-May-28 |
|---|---|---|---|---|---|---|---|---|---|---|---|---|
| 30-Jun-27 | 1.01 | 1.01 | 1.02 | 1.03 | 1.04 | 1.04 | 1.05 | 1.06 | 1.06 | 1.07 | 1.08 | 1.09 |
| 31-Jul-27 | - | 1.01 | 1.01 | 1.02 | 1.03 | 1.04 | 1.04 | 1.05 | 1.06 | 1.06 | 1.07 | 1.08 |
| 31-Aug-27 | - | - | 1.01 | 1.01 | 1.02 | 1.03 | 1.04 | 1.04 | 1.05 | 1.06 | 1.06 | 1.07 |
| 30-Sep-27 | - | - | - | 1.01 | 1.01 | 1.02 | 1.03 | 1.04 | 1.04 | 1.05 | 1.06 | 1.06 |
| 31-Oct-27 | - | - | - | - | 1.01 | 1.01 | 1.02 | 1.03 | 1.04 | 1.04 | 1.05 | 1.06 |
| 30-Nov-27 | - | - | - | - | - | 1.01 | 1.01 | 1.02 | 1.03 | 1.04 | 1.04 | 1.05 |
| 31-Dec-27 | - | - | - | - | - | - | 1.01 | 1.01 | 1.02 | 1.03 | 1.04 | 1.04 |
| 31-Jan-28 | - | - | - | - | - | - | - | 1.01 | 1.01 | 1.02 | 1.03 | 1.04 |
| 29-Feb-28 | - | - | - | - | - | - | - | - | 1.01 | 1.01 | 1.02 | 1.03 |
| 31-Mar-28 | - | - | - | - | - | - | - | - | - | 1.01 | 1.01 | 1.02 |
| 30-Apr-28 | - | - | - | - | - | - | - | - | - | - | 1.01 | 1.01 |
| 31-May-28 | - | - | - | - | - | - | - | - | - | - | - | 1.01 |

*Figure 3.2: The same calculation switched to return the reverse-index step, for testing.*

Instantly, a user or auditor can view the separate calculation blocks and what they do and output. This is significantly more effective than purely reading a formula and attempting to understand the logic.

**How do Financial Actuals get handled**

ERPs, BI, and other accounting software usually have more in common with programmatic approaches to data handling and financial data than traditional financial models. A lot of this has to do with the visibility and variability of a financial forecast that is built and driven from the finance function. When you must be nimble and respond quickly, Excel is usually the best tool. Financial actuals, however, are treated very similarly across systems under and due to accounting standards. Revenues, equity and liabilities have natural credit (negative) balances, expenses and assets have natural debit (positive) balances. If you sum all of the transactions that happened in a particular period, their balance (debits and credits) must equal 0. For the income statement, each account code has an associated location on the income statement. This may be assigned to revenue, COGS, non-operating expenses etc. This structured approach to handing financial actuals makes reporting on the numbers much easier for accountants and data professionals. Instead of a manual approach making a cell reference to a part of the workbook, there is a dimension attribute that is solid and scalable. Balance sheet works the same way, except all of the P&L accounts will get coded to Retained Earnings (potentially with some more specific current year profit logic). Cash flow is usually calculated outside of the financial system but can be compiled for a direct cash flow by simply mapping each GL account to either a category of Operating, Investing,

Financing, Cash and Non-cash. Non-cash would account for anything that directly nets off against itself (like depreciation and accumulated depreciation). Besides this cash movements would equal the net movements in operating, investing and financing.

### Traditional Methods linking and Errors

The traditional method of putting together a financial model involves a build it and link it philosophy. First a modeller will build the financial statements and the associated calculations (like cash in balance sheet equalling ending cash from the cash flow.) The modeller will largely rely on a financial model's balance sheet check to ensure that the model is balancing. This largely works well, but can come undone when the modeller must link multiple sections to the same line, potentially missing a link or a negative sign etc. In each statement the financial modeller (who is not always of a pure accounting background), needs to consider where the flow needs to sit in the income statement, the balance sheet and the cash flow. I note that for income statement items, the balance sheet and cash flow spots are inferred from the net income.

## 4. A Novel Approach

### Mapping of Forecast flows

We take an approach similar to the way we would treat actuals to the problem of mapping forecast flows to the financial statements (CRL method). We create a series of tables that help us to categorise our mapping in a consistent and repeatable manner; one each for income statement, balance sheet and cash flow.

tblIncomeStatementMap

| Income Statement Line | Income Statement Category | Natural balance |
|---|---|---|
| Orignation | Revenue | -1 |
| Trail | Revenue | -1 |
| Broker Commission | Operating Expenses | 1 |
| Broker Trail | Operating Expenses | 1 |
| Marketing | Operating Expenses | 1 |
| Direct Salaries | Operating Expenses | 1 |
| | Net Margin | |
| Salaries | Overheads | 1 |
| IT Expense | Overheads | 1 |
| | EBITDA | |
| Depreciation | Depreciation & Ammortisation | 1 |
| Ammortisation | Depreciation & Ammortisation | 1 |
| | EBIT | |
| Interest Expense | Finance Expense | 1 |
| Interest Income | Interest Income | -1 |
| | EBT | |
| Income Tax Expense | Tax Expense | 1 |
| | Net Income | |
| Overheads | Overheads | 1 |

*Figure 4.1: tblIncomeStatementMap — maps each income statement line to its category and natural balance sign.*

tblBalanceSheetMap

| Balance Sheet Line | Balance Sheet Category | Natural balance |
|---|---|---|
| Cash & cash equivalents | Current assets | 1 |
| Accounts receivable | Current assets | 1 |
| Property plant and equipment | Non-current assets | 1 |
| Prepaid Expenses | Non-current assets | 1 |
| Other intangible assets | Non-current assets | 1 |
| | Assets | 1 |
| Trade and other payables | Current liabilities | -1 |
| Overdraft Facility | Current liabilities | -1 |
| Current tax payable | Current liabilities | -1 |
| Long-term debt | Non-current liabilities | -1 |
| Other non-current laibilities | Non-current liabilities | -1 |
| | Liabilities | -1 |
| Common stock | Equity | -1 |
| Retained Earnings | Equity | -1 |
| Check | Check | -1 |

*Figure 4.2: tblBalanceSheetMap — maps each balance sheet line to its category and natural balance sign.*

tblCashFlowMap

| Cash Flow Line | Cash Flow Category | Ignore for Cash Calcs |
|---|---|---|
| Cash Receipts | Operations | ☐ |
| Cash Payments | Operations | ☐ |
| Prepayments | Operations | ☐ |
| Capex | Investing | ☐ |
| Sale of Trail Book | Investing | ☐ |
| Dividends | Financing | ☐ |
| Loan Drawdown | Financing | ☐ |
| Overdraft Payback (Drawdown) | Financing | ☑ |
| Starting Cash | Starting Cash | ☐ |
| Cash Movements | Cash Movements | ☐ |
| Ending Cash | Ending Cash | ☐ |
| Check | Cash | ☐ |
| Equity Issuance | Financing | ☐ |
| Equity Issuance | Financing | ☐ |

*Figure 4.3: tblCashFlowMap — maps each cash flow line to its category, with a flag for items to ignore in cash calculations.*

We have a mapping table that creates drop down lists which the model will use to assign where they sit in the balance sheet, income statement and cash flow. The signage of each is auto calculated based on modelling and accounting conventions. This forces consistency through the model outputs. Instead of hunting for a negative sign that's causing your model to be out of balance, users understand repeatable and tested model logic, and always know where to go to test their outputs.

*(See Figure A.1 in the Appendix for the full account-map table, tblAcctMap.)*

From there, the structure is built, bringing it into the model becomes the main part. Each workings tab (a tab that has values that can go to the outputs section), has a column that is a named range something along the structure of "*sheet_workings_workings_category*" or "*sheet_cash_workings_category*". This corresponds to an area named range of the same number of rows, that directly references the width of the model, and same length of the category name. All of the workings' categories are VSTACKed together, as with the working areas as well. Wrap each with a filter that the workings category column has a value in it that exists in the account map (CRL) table. This creates "*Workings_Categories*" and "*Workings_Area*". From there, we can access the results from each line, to flow independently into each of the three financial statements.

Income statement:

```
=LET(
 _ISCategory,   XLOOKUP(WorkingsCategory, tblAcctMap[Code],
               tblAcctMap[Income Statement Line], 0),
 _FilterArea,   FILTER(WorkingsArea, _ISCategory=F7),
 _SignReversal, FILTER(IF(XLOOKUP(WorkingsCategory, tblAcctMap[Code],
               tblAcctMap[Sign Reversal])=TRUE, -1, 1), _ISCategory=F7),
 _SignedArea,   _SignReversal * _FilterArea,
 _Result,       IFERROR(BYCOL(_SignedArea, SUM), Timing!$L$11#*0),
 _Result)
```

Balance Sheet category:

```
=LET(
 _BSCategory,   XLOOKUP(WorkingsCategory, tblAcctMap[Code],
               tblAcctMap[Balance Sheet Line], 0),
 _FilterArea,   FILTER(WorkingsArea, _BSCategory=F41),
 _SignReversal, FILTER(IF(XLOOKUP(WorkingsCategory, tblAcctMap[Code],
               tblAcctMap[Sign Reversal])=TRUE, -1, 1), _BSCategory=F41),
 _SignedArea,   _SignReversal * _FilterArea,
 _Result,       IFERROR(BYCOL(_SignedArea, SUM), Timing!$L$11#*0),
 SCAN(0, _Result, LAMBDA(a,b, a+b)))
```

Cash flow:

```
=LET(
 _CFCategory,      XLOOKUP(WorkingsCategory,
```

```
                tblAcctMap[Code], tblAcctMap[Cash Flow Line], 0),
_FilterArea,       FILTER(WorkingsArea, _CFCategory=F75),
_SignReversal,      FILTER(IF(XLOOKUP(WorkingsCategory,
                tblAcctMap[Code], tblAcctMap[Sign Reversal])=TRUE,
                -1, 1), _CFCategory=F75),
_CashFlowReversal,  FILTER(IF(XLOOKUP(WorkingsCategory, tblAcctMap[Code],
                tblAcctMap[Sign Reversal (Cash Flow Only)])=TRUE,
                -1, 1), _CFCategory=F75),
_SignedArea,       _CashFlowReversal * _SignReversal * _FilterArea,
_Result,          IFERROR(BYCOL(_SignedArea, SUM), Timing!$L$11#*0),
_Result)
```

Similar logic, specific mapping. And this allows the logic that drives your model to behave more like a system than a spreadsheet. This method enables a faster, more auditable and trusted modelling process.

**Programmatic Ideas to Standardise Outputs**

One of the things that dynamic arrays unlock is the ability to define a named LAMBDA function that operates on a 2D basis and auto-expands to fit whatever data is in the spreadsheet. Instead of manually maintaining an output block to the current parts of the model and needing to remember to change each time the model shifts, the LAMBDA uses the dimensions of its inputs and spills the answer to an appropriately-sized spill range. This named LAMBDA approach allows for dynamic financial statement outputs in a financial model. When the modeller adds a row of detail to a workings tab and the income statement, balance sheet, and cash flow lines can grow with it. Remove a row and they shrink. The structure of the model reflects the structure of the data, rather than being a fixed grid that the modeller has to maintain by hand.

The trade-off is formatting. Excel doesn't spill formatting the way it spills values, so when an output block change shape the cell borders, fonts, and number formats don't follow. There are two ways to handle this. The first is the heavy use of conditional formatting, where the formatting rules are tied to the data rather than specific cells, and apply themselves to whatever range the LAMBDA happens to fill. The second is a more focussed approach to formatting in general, where you accept that auto-expanding outputs will need to be maintained manually. This is like what has always been required in financial modelling, it just comes with the audit comfort that anything that matches the logic to come into the outputs, will make it there.

There are also actions that can be taken to ensure that the model is working properly, that go beyond the simpler logic in play.  Check cells can be designed to ensure that the values selected in the Workings named ranges are being used in the statements.

**Avoiding Model Circularity**

Model circularity is a common issue in dynamic array financial modelling. It arises when corkscrew calculations depend on outputs from other interrelated parts of the model. Typical examples include revolver or overdraft facilities, cash interest, tax, and other bespoke logic.

When model circularity is introduced in most dynamic array models, iterative calculations are required for the model to calculate properly. This can be very problematic however for models that are of a larger size and significantly increases the chances of errors creeping into the model. Once iterative calculations are on, circular references are harder to find and errors can lie in the model without users or developers knowing.

This is another huge advantage of the CRL approach. It allows the calculations of the model circularity problem to be confined to a separate sheet, for clarity. It also allows the user to quickly pull the cash flow impact out of prior segments of the model, to simplify the model circularity impact within the model.

As a modeller this is great, you can reuse the model circularity schedules where appropriate and then bring in other custom logic as required. As a model auditor it's great too, you have an additional point for checks, and all your most dependant model schedules are in the same place, ready for you to dive in.

**Actuals tie in**

One of the problems that lots of people face in the financial systems are how they bring their actuals in their forecast. Often, traditional modellers are copy pasting system exports in, and more established modellers might connect to APIs from the clients accounting system. As mentioned in the financial actuals section, most good systems will map with their actuals data. This is even more of an argument for the CRL approach to forecasting, using dimensionality from both a forecast and an actuals perspective. This allows for much simpler rollover of forecast logic and can use dynamic array logic to show actuals when its drivers say it's an actuals period, and forecasts when it’s not.

Instead of requiring them to be on a separate tab with manual copy paste logic, tie your forecasts and actuals together with similar methods, that are driven together with table mapping, and only require updates to mapping when things change in your actuals, instead of requiring constant modifications to the financial model for it to continue to be usable each month.

**Audit and Checks**

It’s one thing to build a financial model. It’s another thing for the financial model to be accurate, auditable and audited. It is easy to miss something in this chain. An auditor will check assumptions to ensure they are reasonable, check formulas to ensure that they are working as intended according to their description, and check linking to the model outputs to ensure that everything works as the original modeller intended. Fresh eyes can find errors, ask the right questions, and ensure that the model is as accurate as possible. In a traditional model, a modeller needs to check every line to ensure that the modelling blocks are consistently applied both to the right and downwards. They must ensure that hardcoded numbers aren’t at play, and they need to ensure that the linking to statements is done appropriately, with the right signage applied. This requires a significant level of skill and experience.

Auditing a model with dynamic arrays has a number of similarities to a traditional model, but it requires less checking of cells to ensure consistency across calculation blocks, as such inconsistency is impossible. The auditor can focus their energies on ensuring the logic is right. The CRL approach builds on this again. Instead of needing to understand how each schedule links to the income statement, balance sheet and cash flow, an auditor needs to only understand how the CRL method links all schedules to the statements, which are then applied consistently throughout the model.

Due to the consistent nature of the data brought into outputs with the CRL approach, users can create/utilise audit review sections of the model with ease to help the modeller and auditor to understand the flows of the model. In the worked example above with regards to marketing expense, an auditor can view the audit tab, filter for marketing, and see its impact on the income statement, balance sheet, cash flow and balance sheet flows (just the movements, as opposed to continuing balance).

The Model-Audit tab for this worked example (filtered for Marketing) is shown at full width in the Appendix, Figure A.2.

Users can use these filters to understand what is driving different parts of each of the financial statements. And each section comes with a hyperlink that allows the user to directly view the part of the model that they are interested in. The audit tab using the CRL method serves as a diagnostic tool for the modeller to help them to understand their model better.

The checks section of a financial model is extremely important in ensuring the trust-worthiness of the model and its outputs. There will be checks such as balance sheet balancing to zero, but other types of calculations when you can ensure parts of the model flow together properly are important too. Checks also serve as an important diagnostic tool, because they can be displayed throughout the model, so that the modeller always has the comfort of knowing that if one of the checks is wrong, it will flag loudly to the user, who can check the model. Each section of the checks can flag individually, so the user can be sure which part of the model is not functioning as intended.

A key benefit of the CRL method, in providing a consistent methodology for linking schedules to the output, is that checks can be applied to ensure that the forecasted flows that are intended for the output are actually present. This is a sharp contrast to the traditional approach, which requires manual review or a 'gut instinct' feeling to tell you that a particular number isn't quite right.

**Advanced Formula Environment and the Named LAMBDA**

Whilst dynamic arrays and the **LET** function had delivered significant value in structuring spreadsheet solutions, it was **LAMBDA** that truly revolutionised Excel, turning its formula language into a Turing complete programming environment.

Dynamic arrays allow the financial modeller to express cash flows and the associated balances as functions of time, rather than as isolated values. **LET** can take complex nested formulas and lay them out as a sequence of statements, the intermediate results being assigned to variables with meaningful names.

**LAMBDA** builds upon the **LET** functionality, but instead applies local names to external references that are listed as arguments to the **LAMBDA** function. The name the user choses for the function itself should describe its purpose. Typically **LAMBDA** functions are developed within the Advanced Formula Editor which supports both structured layout of the code and documentation provided in the form of inline comments. Some traditionalist modellers may view this with suspicion as a move towards becoming a black box. The reality should be somewhat different; the AFE is serving the same role as the existing formula bar and provides open access to the functionality of the model for both developers and auditors.

A named **LAMBDA** operates in very similar manner to any native Excel function. You are able to pass parameters into it and return a calculated output. This can allow for reusable functions across your workbook. With integration to Github gists, you can even maintain your formulas from the cloud, and pass updates down centrally, and with speed.

For regular formulas, the benefit is a structured environment that wrap the complexity into a place that the developers know where to access it. This is consistent with other programming. For formulas that may be repeated throughout the workbooks or a portfolio of models, the benefits grow exponentially. Named **LAMBDA** functions are able to standardise the regular functionality that appears, so that changes, updates can be applied centrally with substantially less work required. Your 'plumbing' can provide a consistent standard that the whole firm can learn from and utilise. Instead of the traditionally accepted logic that each modeller develops in the best way they can within time constraints, the 'plumbing' of the flows of data is already available, to be accessed and quickly reconfigured instead of built from scratch.  This will allow less experienced modellers to model at the level of their more experienced team members, by taking advantage of prebuilt LAMBDAs.

### Teaching AI your modelling Structure

Having a consistent modelling format is great for AI, because it allows you to build skills that accelerate your financial modelling. Now, instead of AI doing something different every time, it operates within your guard rails and your modelling structure. And when you develop a financial modelling system like this, the AI doesn't need to perform the linking logic and thinking that causes most AI models to go out of balance in traditional 3-statement modelling. It models, effectively and within a system designed to stop it from making errors.

### Financial Modelling with Business Intelligence

Having a financial model that can integrate well with your business intelligence reporting is important, because it can help to illustrate the long-term planning to a larger audience. Traditionally, it has been possible to bring financial models into business intelligence tools, by using a named range and a big unpivot, but being able to access the underlying data generated, required significant manual mapping and logic for each PBI implementation.

With a consistent modelling structure, centred around a mapping logic illustrated in the prior sections, financial models are able to be directly read by business intelligence systems and allow users to access the data at the level of the forecast. This is important if you have for example revenue models that calculate down at a detailed level, giving the users the ability to access this granularity for drill throughs and subsequent analysis. It also enables much more repeatable logic when linking a financial model to a reporting system.

It also allows the bridge between actuals and forecasts to become significantly easier because that data is already there and available, and crucially, uses the same labels.

## 5. A worked example

To illustrate how easy it is to add a new line to a model built this way, take a simple worked example. Say we want to add a monthly marketing spend to an equity release model. In a traditional model this would mean building the calculation, then linking it by hand into the income statement, the cash flow statement and the net income would be pulled back into another schedule to calculate retained earnings which gets linked back again to the balance sheet. With the CRL approach, we map it once and let the structure do the rest. The whole process is six steps, and most of them are single clicks.

*tblAcctMap*

| Code | Income Statement Line | Balance Sheet Line | Cash Flow Line |
|---|---|---|---|
| Marketing | Marketing | Retained Earnings | Cash Payments |

*Figure 16.1: A new "Marketing" code added to tblAcctMap, routing it to the Income Statement (Marketing), Balance Sheet (Retained Earnings) and Cash Flow (Cash Payments).*

We start in the account map. We add a “Marketing” code and tell it where the movement belongs in each statement — Marketing in the income statement, Retained Earnings in the balance sheet, and Cash Payments in the cash flow. The signage is handled for us by the conventions already built into the map. That is the only place we have to make a routing decision, and we only make it once.

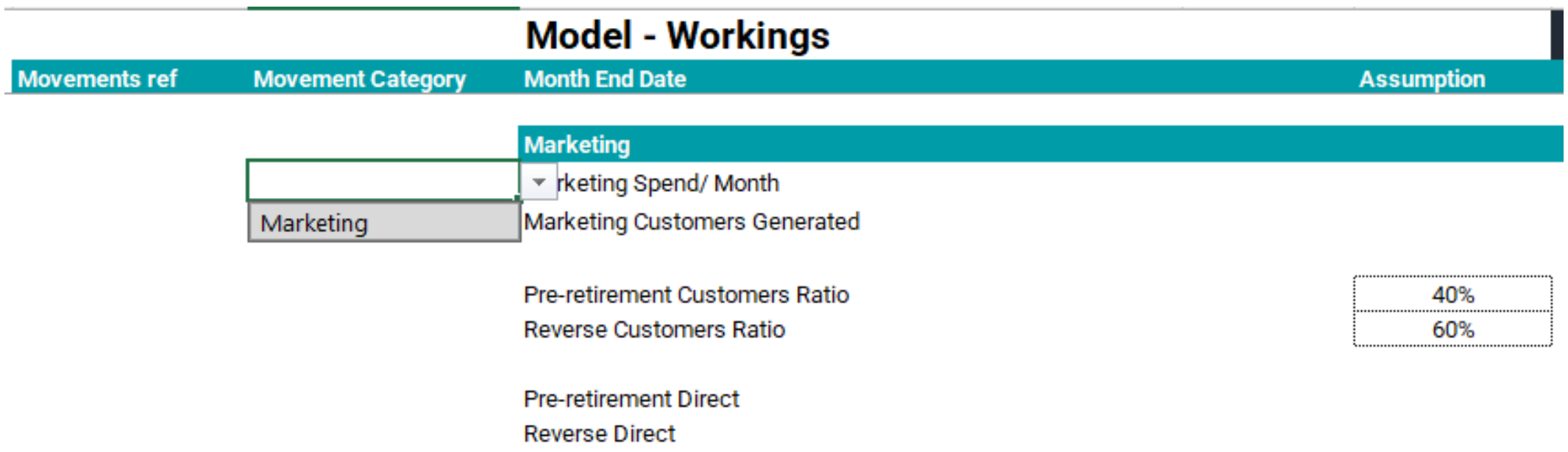

*Figure 16.2: Selecting “Marketing” from the Movement Category drop-down against the working line on the Workings tab.*

From there we go to the working line itself and pick “Marketing” from the Movement Category drop-down. That drop-down is driven by the same account map, so the only options it offers are ones that already have a home in the statements. There is no linking to do and no formula to write. The moment the category is selected, the line is live across the whole model.

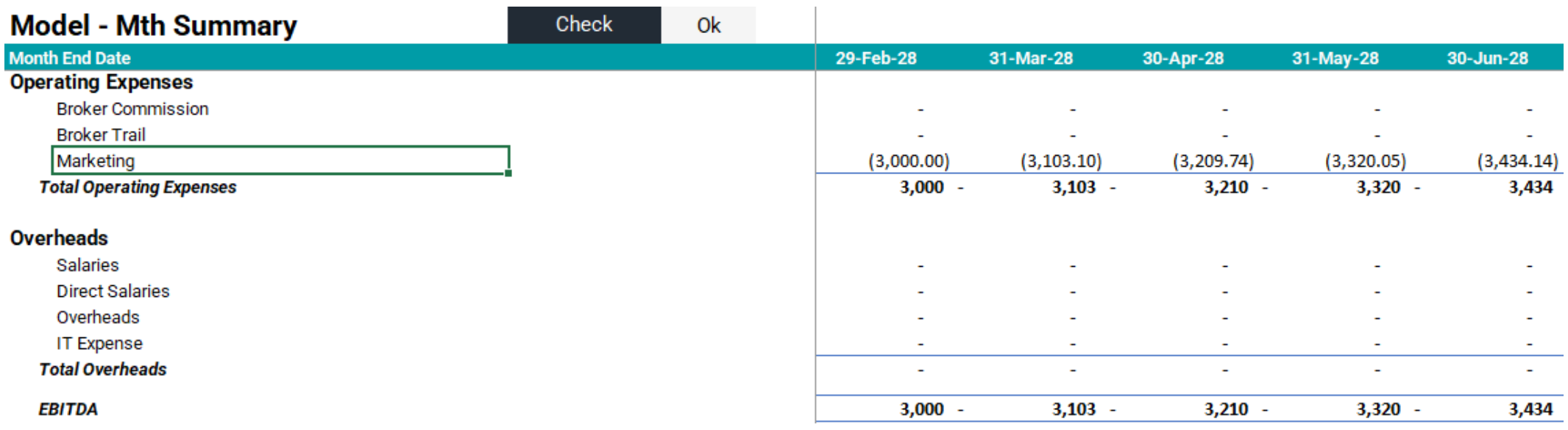

*Figure 16.3: The Income Statement, with the Marketing expense flowing automatically into Operating Expenses.*

And that is it. With nothing more than a code and a drop-down selection, the marketing spend now flows into all three statements at once, in the right place and with the right sign.

**Model - Mth Summary** | Check | Ok

| Month End Date | 29-Feb-28 | 31-Mar-28 | 30-Apr-28 | 31-May-28 | 30-Jun-28 |
|---|---|---|---|---|---|
| **Cash Flow** | | | | | |
| **Operations** | | | | | |
| Cash Receipts | - | - | - | - | - |
| Cash Payments | (3,000) | (3,103) | (3,210) | (3,320) | (3,434) |
| Wages | - | - | - | - | - |
| Tax | - | - | - | - | - |
| Prepayments | - | - | - | - | - |
| ***Total Operations*** | **(3,000)** | **(3,103)** | **(3,210)** | **(3,320)** | **(3,434)** |
| **Investing** | | | | | |
| Capex | - | - | - | - | - |
| Sale of Trail Book | - | - | - | - | - |
| ***Total Investing*** | - | - | - | - | - |
| **Financing** | | | | | |
| Dividends | - | - | - | - | - |
| Overdraft Payback (Drawdown) | - | - | - | - | - |
| Equity Issuance | | | | | |
| Loan Drawdown | - | - | - | - | - |
| ***Total Financing*** | - | - | - | - | - |
| Starting Cash | - | (3,000) | (6,103) | (9,313) | (12,633) |
| Cash Movements | (3,000) | (3,103) | (3,210) | (3,320) | (3,434) |
| Ending Cash | (3,000) | (6,103) | (9,313) | (12,633) | (16,067) |

*Figure 16.4: The Cash Flow, with the same movement appearing as a Cash Payment under Operations.*

**Model - Mth Summary** | Check | Ok

| Month End Date | 29-Feb-28 | 31-Mar-28 | 30-Apr-28 | 31-May-28 |
|---|---|---|---|---|
| **Balance Sheet** | | | | |
| **Assets** | | | | |
| Cash & cash equivalents | (3,000.00) | (6,103.10) | (9,312.84) | (12,632.88) |
| Accounts receivable | - | - | - | - |
| ***Total Current assets*** | **3,000** - | **6,103** - | **9,313** - | **12,633** - |
| Property plant and equipment | - | - | - | - |
| Deferred tax asset | - | - | - | - |
| ***Total Non-current assets*** | - | - | - | - |
| ***Total Assets*** | **3,000** - | **6,103** - | **9,313** - | **12,633** - |
| **Liabilities** | | | | |
| Trade and other payables | - | - | - | - |
| Current tax payable | - | - | - | - |
| ***Total Current liabilities*** | - | - | - | - |
| Long-term debt | - | - | - | - |
| Other non-current laibilities | - | - | - | - |
| ***Total Non-current liabilities*** | - | - | - | - |
| ***Total Liabilities*** | - | - | - | - |
| **Equity** | | | | |
| Common stock | - | - | - | - |
| Retained Earnings | (3,000.00) | (6,103.10) | (9,312.84) | (12,632.88) |
| ***Total Equity*** | **3,000** - | **6,103** - | **9,313** - | **12,633** - |

*Figure 16.5: The Balance Sheet, where the movement accumulates into Retained Earnings and Cash.*

**Model - Audit** | Check | Ok

| Code | Statement | Type | Category | Line | Detail | Cell | Units | 30-Sep-28 | 31-Oct-28 | 30-Nov-28 | 31-Dec-28 |
|---|---|---|---|---|---|---|---|---|---|---|---|
| | | | | | | | | (38,495.44) | (42,818.37) | (47,289.87) | (51,915.04) |
| Marketing | Income Statement | IS | Operating Expenses | Marketing | Marketing Spend/ Month | Workings!L5 | $ | (3,800.50) | (3,931.11) | (4,066.21) | (4,205.95) |
| Marketing | Balance Sheet | BS Flow | Equity | Retained Earnings | Marketing Spend/ Month | Workings!L5 | $ | (3,800.50) | (3,931.11) | (4,066.21) | (4,205.95) |
| Marketing | Balance Sheet | BS Balance | Equity | Retained Earnings | Marketing Spend/ Month | Workings!L5 | $ | (27,093.93) | (31,025.04) | (35,091.25) | (39,297.20) |
| Marketing | Cash Flow | CF | Operations | Cash Payments | Marketing Spend/ Month | Workings!L5 | $ | (3,800.50) | (3,931.11) | (4,066.21) | (4,205.95) |

*Figure 16.6: The Audit tab, showing the Marketing code's routing to all three statements with hyperlinks back to the source working line.*

The audit tab closes the loop. It shows the new code routed to every statement it touches, with the sign applied and a hyperlink back to the working line that drove it. For an auditor, that is the whole story of the change on one screen — what was added, where it went, and why. A new line that would have taken careful linking and checking in a traditional model is here a code and a click, and the model stays auditable the entire way through.

## 6. Limitations

Whilst the CRL method has some significant advantages over more traditional methodologies, it is worth noting some potential drawbacks. One of the main reasons that people use Excel is because it is the ultimate fall back in terms of flexibility, to allow users to do what they need to do, and it also speaks in a language that regular users can understand. The CRL method utilises more advanced techniques that less advanced users may not be able to modify as freely. This includes named formulas or **LAMBDA's** using the advanced formula environment. There are a limited number of modellers and auditors with the skills to read models using this functionality. The use of these named **LAMBDA's** can be limited to the repeated and consistent parts of a model that are either reused in models or used throughout a model. Maintaining a named **LAMBDA** does mean that modification of the logic of the formula is able to be controlled at a global level, and changes do not need to be made in every individual formula block.

Using dynamic arrays requires a spill area for values to display into. It is highly possible for users to modify assumptions which can cause the spill area to expand and no longer have enough space. When this happens, a #SPILL error will occur. Whilst it is easy to insert rows to allow more space, this is a potential failure point, especially in users not experienced with dynamic arrays. In more traditional models, the types of changes, especially to 2D structures, would require spreadsheet developer/modeller involvement, which ensures that the resized array is appropriately adjusted. One method dynamic array modellers are using is to allow blank space below their calculation blocks that are spilling, to reduce the risk and annoyance of #SPILL errors.

Computation speed in dynamic arrays is generally faster than the traditional cell approach. Where it can be slower is when recursion is required to achieve a calculation. The CRL approach largely attempts to avoid recursion where possible, and if model circularity starts limiting the dynamic array modelling process, traditional cells can be used on the Cash tab as an acceptable practice. The other potential computational inefficiency is that the CRL method requires each line item to view a named range containing all calculation blocks and filter for only the appropriate parts. This will happen for every line item within the model. In very large models, that individual line may require looking at 1000's of rows to find the rows intended for that part of the statement. This can cause inefficiencies and a slower model than all direct references. Direct references require more linking, intermediary calculations and limit the available use of dynamic arrays due to model circularity. This is a trade-off that users with very large models will have to consider.

Dynamic arrays and by extension, the CRL method are built on the most recent version of Excel, within Microsoft 365. Users that do not have Microsoft 365 will see errors when they attempt to open a workbook containing dynamic array formulas.

## 7. Conclusion

The shift to dynamic arrays in Excel has been more than a feature update. It's a genuine change in how financial models can be built, audited, and trusted. The traditional fill-down-and-right approach can be effective, but it carries a tax of thousands of formula cells, fragility, and silent errors are much harder to catch. Dynamic arrays, **LET**, **LAMBDA**, and the array-shaping functions give us the tools to replace that approach with something closer to a system than a spreadsheet.

The CRL approach is the thread that ties this together. Instead of hand-linking each workings line into the income statement, balance sheet and cash flow, we categorise once and let the structure do the routing. Signage is consistent, the logic is repeatable, and when something is out of balance, users know exactly where to look. It also provides us a clean place to confine model circularity, a natural bridge between forecast and actuals, and much stronger auditability.

With AI becoming increasingly prevalent and starting to legitimately seep into the financial modelling space, guard rails and consistency are becoming increasingly important. This consistent modelling structure allows AI to be useful in a model rather than dangerous, because the AI operates within guard rails instead of inventing its own linking logic each time. The same structure is what lets business intelligence tools read the forecast at the level of detail it was built at, using the same labels as the actuals. Dynamic arrays opened the door to more consistent and structured models. The CRL approach takes advantage of that and optimises for speed and efficiency. As financial modellers, it's on us to build in the most auditable way the tools now allow.

## 8. Appendix

*Figure A.1: tblAcctMap — the master account-map table routing every Workings code to its Income Statement, Balance Sheet and Cash Flow line, with sign-reversal flags.*

tblAcctMap

| Code | Income Statement Line | Balance Sheet Line | Cash Flow Line | Income Statement Category | Balance Sheet Category | Cash Flow Category | Sign Reversal | Sign Reversal (Cash Flow Only) |
|---|---|---|---|---|---|---|---|---|
| Marketing | Marketing | Retained Earnings | Cash Payments | Operating Expenses | Equity | Operations | ☑ | ☐ |
| FwdTrailRevenue | Trail | Retained Earnings | Cash Receipts | Revenue | Equity | Operations | ☐ | ☐ |
| RevTrailRevenue | Trail | Retained Earnings | Cash Receipts | Revenue | Equity | Operations | ☐ | ☐ |
| FwdTrailBrokerExp | Broker Trail | Retained Earnings | Cash Payments | Operating Expenses | Equity | Operations | ☑ | ☐ |
| RevTrailBrokerExp | Broker Trail | Retained Earnings | Cash Payments | Operating Expenses | Equity | Operations | ☑ | ☐ |
| FwdOriginationRevenue | Orignation | Retained Earnings | Cash Receipts | Revenue | Equity | Operations | ☐ | ☐ |
| RevOriginationRevenue | Orignation | Retained Earnings | Cash Receipts | Revenue | Equity | Operations | ☐ | ☐ |
| FwdOriginationBrokerExp | Broker Commission | Retained Earnings | Cash Payments | Operating Expenses | Equity | Operations | ☑ | ☐ |
| RevOriginationBrokerExp | Broker Commission | Retained Earnings | Cash Payments | Operating Expenses | Equity | Operations | ☑ | ☐ |
| DirectSalary | Direct Salaries | Retained Earnings | Cash Payments | Operating Expenses | Equity | Operations | ☑ | ☐ |
| IndirectSalary | Salaries | Retained Earnings | Cash Payments | Overheads | Equity | Operations | ☑ | ☐ |
| TechCosts | IT Expense | Retained Earnings | Cash Payments | Overheads | Equity | Operations | ☑ | ☐ |
| PrepaidExpenses | | Prepaid Expenses | Prepayments | | Non-current assets | Operations | ☐ | ☑ |
| Interest Expense | Interest Expense | Retained Earnings | Cash Payments | Finance Expense | Equity | Operations | ☑ | ☐ |
| Interest Income | Interest Income | Retained Earnings | Cash Receipts | Interest Income | Equity | Operations | ☐ | ☐ |
| Overdraft Movements | | Overdraft Facility | Overdraft Payback (Drawdown) | | Current liabilities | Financing | ☐ | ☐ |
| OverheadCosts | Overheads | Retained Earnings | Cash Payments | Overheads | Equity | Operations | ☑ | ☐ |
| PrepaidOverheads | | Prepaid Expenses | Prepayments | | Non-current assets | Operations | ☐ | ☑ |
| Laptop_Capex | | Property plant and equipment | Capex | | Non-current assets | Investing | ☐ | ☑ |
| Laptop_Depn | Depreciation | Property plant and equipment | Cash Payments | Depreciation & Ammortisation | Non-current assets | Operations | ☑ | ☑ |
| Desk_Capex | | Property plant and equipment | Capex | | Non-current assets | Investing | ☐ | ☑ |
| Depn | Depreciation | Property plant and equipment | Cash Payments | Depreciation & Ammortisation | Non-current assets | Operations | ☑ | ☑ |
| Uniforms | Overheads | Retained Earnings | Cash Payments | Overheads | Equity | Operations | ☑ | ☐ |
| SeedEquity | | Common stock | Equity Issuance | | Equity | Financing | ☐ | ☐ |
| BusinessLoanDrawdown | | Long-term debt | Loan Drawdown | | Non-current liabilities | Financing | ☐ | ☐ |
| BusinessLoanPrincipal | | Long-term debt | Loan Drawdown | | Non-current liabilities | Financing | ☐ | ☐ |
| BusinessLoanInterest | Interest Expense | Retained Earnings | Cash Payments | Finance Expense | Equity | Operations | ☑ | ☐ |
| Accumulated Depn | | Property plant and equipment | Cash Payments | | Non-current assets | Operations | ☐ | ☑ |

*Figure A.2: The Model-Audit tab, filtered for Marketing — showing its flow into the income statement, balance sheet and cash flow.*

**Model - Audit**

| | | | Check | Ok | | | | (18,942.06) | (22,593.02) | (26,369.45) | (30,275.67) |
|---|---|---|---|---|---|---|---|---|---|---|---|
| **Code** | **Statement** | **Type** | **Category** | **Line** | **Detail** | **Cell** | **Units** | **30-Apr-28** | **31-May-28** | **30-Jun-28** | **31-Jul-28** |
| Marketing | Income Statemen | IS | Operating Expenses | Marketing | Marketing Spend/ Month | Workings!L5 | $ | (3,209.74) | (3,320.05) | (3,434.14) | (3,552.16) |
| Marketing | Balance Sheet | BS Flow | Equity | Retained Earnings | Marketing Spend/ Month | Workings!L5 | $ | (3,209.74) | (3,320.05) | (3,434.14) | (3,552.16) |
| Marketing | Balance Sheet | BS Balance | Equity | Retained Earnings | Marketing Spend/ Month | Workings!L5 | $ | (9,312.84) | (12,632.88) | (16,067.03) | (19,619.19) |
| Marketing | Cash Flow | CF | Operations | Cash Payments | Marketing Spend/ Month | Workings!L5 | $ | (3,209.74) | (3,320.05) | (3,434.14) | (3,552.16) |